# Optimal axonal and dendritic branching strategies during the development of neural circuitry

Dmitry Tsigankov and Alexei Koulakov,
Cold Spring Harbor Laboratory, 1 Bungtown Rd, Cold Spring Harbor, NY, 11724, USA

## Abstract

In developing brain, axons and dendrites are capable of connecting to each other with high precision. Recent advances in imaging have allowed for the monitoring of axonal, dendritic, and synapse dynamics *in vivo*. It is observed that the majority of axonal and dendritic branches are formed 'in error', only to be retracted later. The functional significance of the overproduction of branches is not clear. In this study, we use a computational model to investigate the speed and efficiency of different branching strategies. We show that branching itself allows for substantial acceleration in the identification of appropriate targets through the use of a parallel search. We also show that the formation of new branches in the vicinity of existing synapses leads to the formation of target connectivity with a decreased number of erroneous branches. This finding allows us to explain the high correlation between the branch points and synapses observed in the *Xenopus laevis* retinotectal system. We also suggest that the most efficient branching rule is different for axons and dendrites. The optimal axonal strategy is to form new branches in the vicinity of existing synapses, whereas the optimal rule for dendrites is to form new branches preferentially in the vicinity of synapses with correlated pre- and postsynaptic electric activity. Thus, our studies suggest that the developing neural system employs a set of sophisticated computational strategies that facilitate the formation of required circuitry, so that it may proceed in the fastest and most frugal way.

## Introduction

Neural development is a dynamic process that leads to the establishment of precise connectivity (Ruthazer and Cline, 2004). *In vivo* time lapse imaging has shown that the formation of axonal and dendritic arbors involves the simultaneous creation and elimination of neuronal branches and synapses (Alsina et al., 2001; Niell et al., 2004; Haas et al., 2006; Meyer and Smith, 2006; Ruthazer et al., 2006; Sanchez et al., 2006). The high rate of branch turnover results in the formation of a number of branches that substantially exceeds the number maintained in the mature brain (Rajan et al., 1999; Meyer and Smith, 2006). These observations suggest that a form of 'trial-and-error' search algorithm is implemented by axons and dendrites (Hua and Smith, 2004).

The branching of axons and dendrites depends upon the synapses they form. First, branch survival depends on the presence and strength of the synapses it bears (Niell et al., 2004; Meyer and Smith, 2006; Ruthazer et al., 2006). Second, a spatial bias of the locations of branch points towards synapses has been reported (Alsina et al., 2001; Meyer and Smith, 2006), suggesting that new branches are formed preferentially in the vicinity of synapses. Finally, it has been shown that the rates of branch additions and retractions are affected by neuronal electric activity. These rates increase for retinal axons and decrease for tectal dendrites after an NMDAR antagonist is applied to a developing *Xenopus laevis* retinotectal system (Rajan et al., 1999; Sin et al., 2002). The branching rules, therefore, are different for axons and dendrites. The functional significance of the asymmetry between axons and dendrites is not clear.

Here, we theoretically investigate the role of branching in the formation of neural connectivity. We ask three questions stemming from the experimental findings mentioned above. First, we ask: what is the functional significance of branching, from the standpoint of neural development? Second, we ask why axons and dendrites preferentially form branches in the vicinity of synapses. Third, we address the asymmetry in the branching rules between axons and dendrites that has been revealed in experiments on NMDA receptor blockade. To answer these questions, we have developed a computational model that allows us to compare different branching strategies, based upon the speed of development of target circuitry and the number of 'erroneous' branches formed. We show that three prominent features of axon and dendrite dynamics can be viewed as evolutionary adaptations that save time and minimize the number of errors. We propose experimental tests that can differentiate the various branching strategies used by axons and dendrites.

## Results

**Formation of retinotectal connectivity is influenced by several factors**

The projections from the retina to optic tectum often are used as a model system to study the development of neural circuitry. While establishing this projection, the axons of retinal ganglion cells (RGC) arrive at the optic tectum and make topographically-ordered connections with dendrites in the target. This implies that every two axons of neighboring retinal ganglion cells terminate at proximal tectal dendrites. This form of connectivity often is called a 'topographic map'.

Several factors contribute to the formation of topographic maps. A set of chemical labels is thought to encode coordinates in the retina and tectum (McLaughlin and O'Leary, 2005). Thus, the nasal-temporal (NT) axis in the retina is encoded by the graded expression of EphA receptor tyrosine kinases on RGC axons (Flanagan and Vanderhaeghen, 1998). The recipient anterior-posterior coordinate in the tectum is established by graded expression of ephrin-A, which can bind to and activate EphA receptors, and transmit to RGC axons information about their position in the target. A similar chemical labeling system, involving an EphB/ephrin-B receptor/ligand pair, exists for the mapping of the dorso-ventral (DV) axis of the retina to the medial-lateral (ML) direction of the optic tectum. The two approximately perpendicular expression profiles appear to be in place to bias axonal branching in the direction of the correct termination site (Lemke and Reber, 2005).

The precision of axonal projections is further enhanced through mechanisms based upon correlated neural activity (Ruthazer and Cline, 2004). Due to correlations in the visual stimuli or the presence of retinal waves during development, electrical activity is similar in neighboring RGC axons in the retina (McLaughlin et al., 2003). Correlated activity, therefore, provides additional information about axonal relative positions in retina, and contributes to the precision of topographic projection (McLaughlin et al., 2003; Pfeiffenberger et al., 2005). Finally, competition between axons in the target is thought to be an important factor in the formation of the map (Hua et al., 2005). The interplay of chemo-specificity, activity-dependent factors, and competition results in the formation of connectivity that sometimes can achieve single-cell precision (Hamos et al., 1987).

Precise connectivity requires spatial overlap between an axonal arbor and the arbors of appropriate dendrites. This is because the synapses can be made only between segments of axonal and dendritic branches that are in close proximity; i.e. have the potential to form connectivity (Stepanyants and Chklovskii, 2005). Thus, before appropriate axons and dendrites are connected, they must solve the search problem, which implies that axons have to arrive in the area of appropriate dendrites. This task is achieved by creating and eliminating new axon and dendrite branches (Alsina et al., 2001; Ruthazer et al., 2003; Meyer and Smith, 2006). The exact rules by which axonal and dendritic branching occurs and their functional significance are not known. Here, we identify the axonal and dendritic branching rules that implement the optimal search strategy, based upon the conservation of material and time.

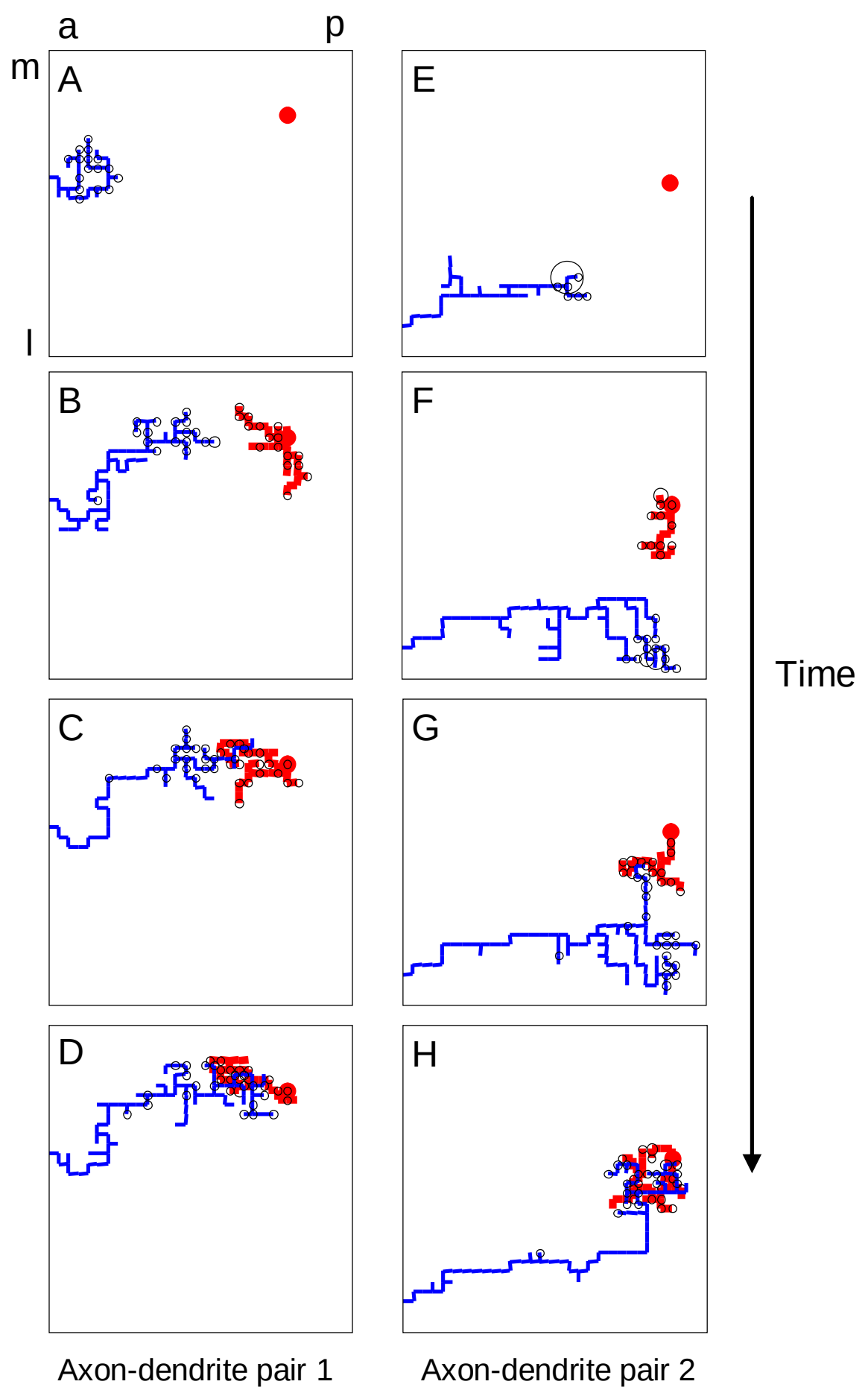

Figure 1: The structure of developing axons and dendrites *in silico* as a function of time. (A-D) The evolution of the axonal (blue) and dendritic (red) arbor in the model is achieved by creating and eliminating new branches and retracting and elongating existing ones. A particular pair of axon and dendrite is shown out of 900 simultaneously evolving axonal and dendritic arbors. The shown dendrite is the main recipient of the synaptic connections (black circles) for the shown axon, once topography is established. (E-H) The evolution of another pair of axon and dendrite is shown for the same simulation.

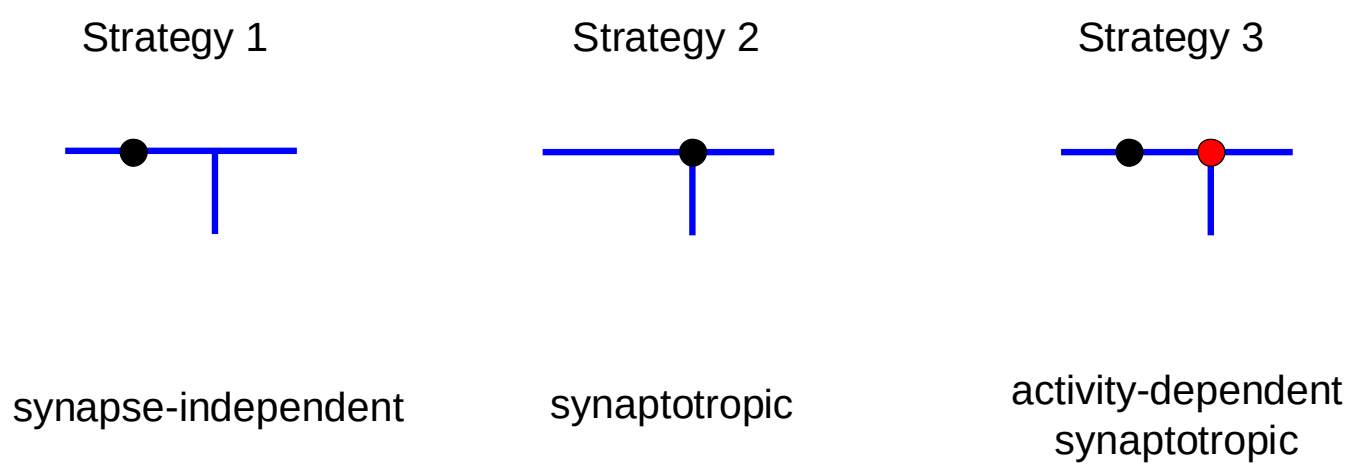

Figure 2: Different branching strategies available to axons and dendrites. Strategy 1: a new branch point on the arbor (blue) can be formed anywhere, independent of the location of synapses (black circle). Strategy 2: new branch points are formed preferentially in the vicinity of existing synapses. Strategy 3: new branch points are formed preferentially in the vicinity of synapses with correlated pre- and postsynaptic activity only (red circle).

### Branching allows for faster formation of target connectivity

To compare various search strategies, we have developed a computational model for the stochastic growth of axonal and dendritic arbors. This model describes the behavior of RGC axons that form synapses with a matching set of tectal dendrites (Figure 1). Both axons and dendrites can create, eliminate, extend, and retract their branches. In addition, an axon and a dendrite with overlapping arbors can form a new synapse or eliminate the existing one. All these events occur stochastically, with probabilities biased towards the formation of a topographic map. A conventional method to describe such a bias is to introduce an energy function (Fraser and Perkel, 1990; Koulakov and Tsigankov, 2004; Tsigankov and Koulakov, 2006). With this approach, the stochastic events of creation and elimination of new branches and synapses are biased in the direction of an overall decrease in energy function. The energy function includes both contributions from the binding and activation of chemical labels, such as Eph receptors, and the contribution arising from the correlations in electric activity that exist between retinal axons (see Methods for details). The exact form of the energy function defines both the dynamics of the arbor formation and the structure of the ultimately established connectivity.

Using this approach, we investigated different branching strategies available to axons and dendrites. One possibility is that formation of new branches occurs everywhere on the arbor with the same probability, independent of the locations of synapses. We call this type of branching strategy *synapse-independent* or *Strategy 1* (Figure 2). Another option is to preferentially form new branches in the vicinity of existing synapses. This strategy is called *synaptotropic* or *Strategy 2*. Finally, we considered the possibility that branches are formed preferentially in the vicinity of synapses with correlated pre- and postsynaptic activity. This form of branching rules is called *activity-dependent* or *Strategy 3*. To implement these branching rules, we introduced the cost of the formation of a branch point that differs for the different strategies. This cost was included in the cost function, as described in Methods. We show below that every branching strategy is capable of producing the required connectivity, but their efficiencies differ.

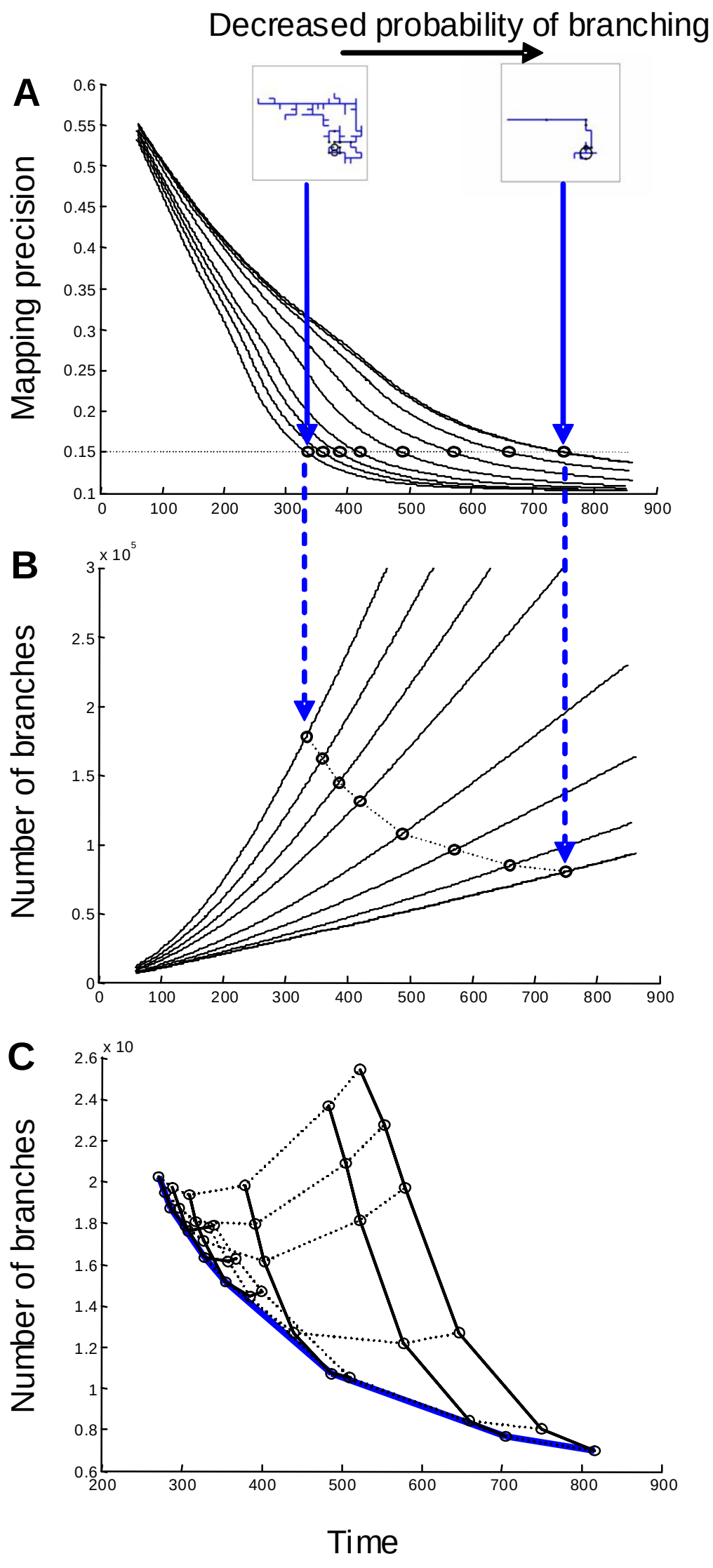

Figure 3: The influence of branching on time and material cost. (A) The time-dependence of mapping precision is shown for simulations involving different probabilities of axonal branching when both axons and dendrites use branching Strategy 1. When the probability of branching is low, axons have very few branches; meanwhile, when the probability is high, the axonal arbors are complex (inset). Mapping with the same degree of precision is established faster when the probability of branching is higher. (B) The time-dependence of the number of branches formed is shown for the same set of simulations, as in (A). Each circle in (B) corresponds to a circle in (A) and shows the time and the number of branches formed when a 15% level of mapping precision is reached. A mapping precision level of 15% implies that the standard deviation of synapse location is 15% of the map size. (C) A set of curves similar to that shown in (B) is obtained when both axonal and dendritic branching probabilities are varied. Points corresponding to the same axon/dendrite branching probabilities are connected by solid/dashed lines. The lower boundary of the collection of these points (blue line in C) gives the performance boundary for this combination of branching strategies used by axons and dendrites.

As measures of the efficiency of different branching strategies, we used the time and the total number of branches formed (dendritic and axonal) that are required to achieve target mapping precision. We propose that a more efficient developmental mechanism should allow for the formation of required connectivity using less physical time and less material for creating and elongating neuronal branches. These two separate criteria are not independent and cannot be minimized simultaneously. In fact, we show that there is a trade-off between time and the number of branches: if connectivity is formed faster, it uses a greater number of branches, and *vice versa*.

To illustrate the trade-off between time and the number of branches formed, we consider the case in which both axons and dendrites implement synapse-independent branching (Strategy 1). One of the parameters that can be varied in the model is the probability of forming a new branch point on an axon. If this probability is small, axon arbors have a simple structure with few branch points (see inset on Figure 3A). Nevertheless, the mapping precision is improved over time and always can reach the target value (Figure 3A), even if virtually no branches are formed. If the branching probability is increased, the arbor structure becomes more complex with more branch tips. This results in a faster convergence of map precision, because multiple branches are searching for the correct partners in parallel. At the same time, higher branching frequency results in a greater number of branches formed by the time connectivity with the required precision is established (Figure 3B). Thus, faster convergence of the map can be accomplished by forming a larger number of branches, implying a trade-off between the time of development and the amount of material used. These findings suggest a possible functional role for axonal branching, as an effective parallel search algorithm that allows for the conservation of time during development.

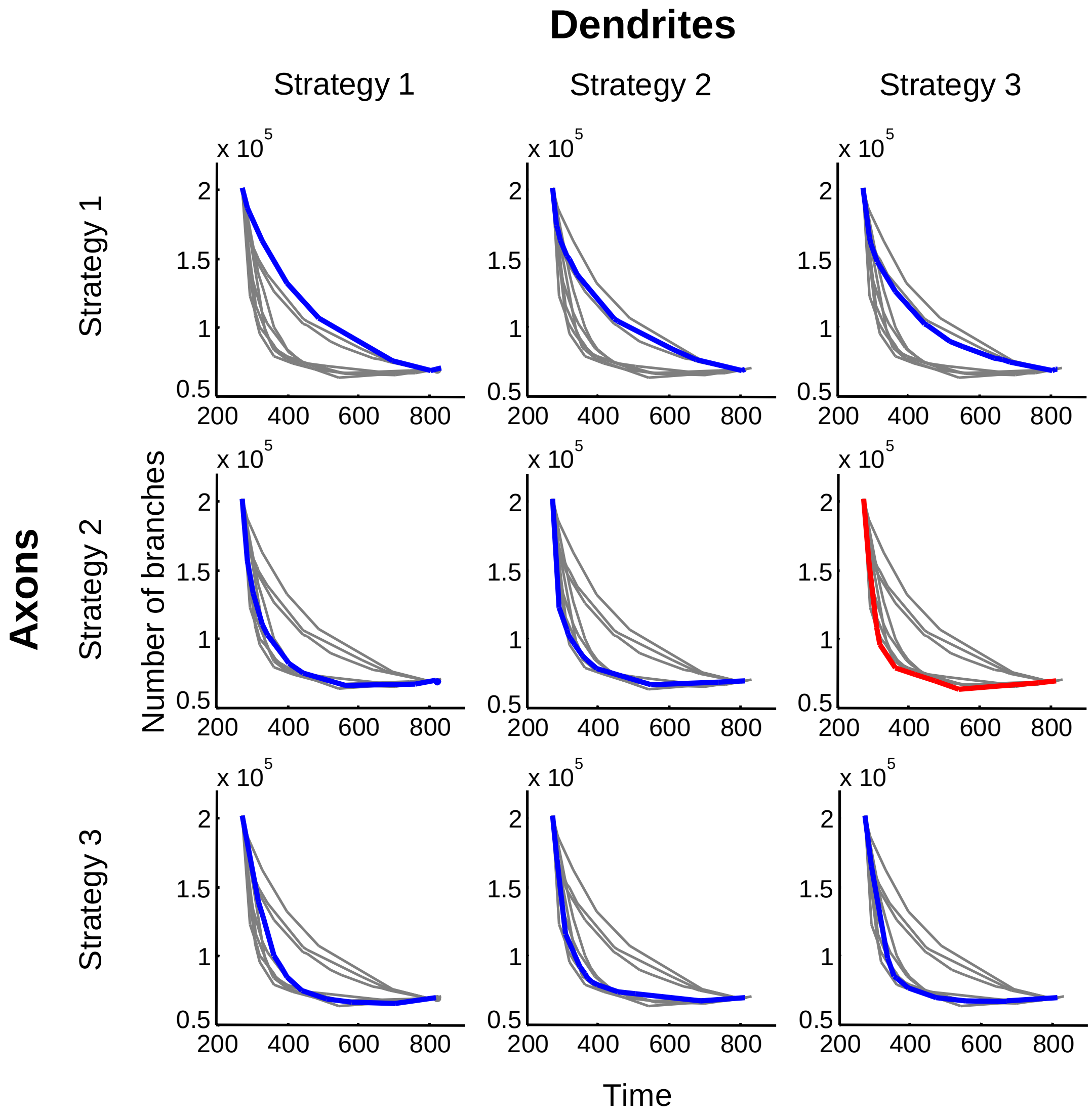

Figure 4: Performance boundaries obtained for 9 combinations of branching strategies used by axons and dendrites. The performance boundary depicts the minimal number of branches that are required to establish connectivity with a given level of precision after a given length of time. Alternatively, it provides the minimal time required to achieve the configuration with a given level of precision using a predetermined number of branches. In each panel, the colored curve (blue or red) represents the boundary for a specified combination of branching strategies. The boundaries for other combinations are shown in grey, for comparison. The red curve depicts the optimal performance boundary that corresponds to the combination of Strategies 2 and 3, used by axons and dendrites, respectively.

**Synaptotropic branching minimizes the total number of branches formed**

We next optimized the total number of branches formed for varying frequencies of both axonal and dendritic branching if they use synapse-independent branching Strategy 1. To this end, we obtained a series of curves similar to those shown in Figure 3B for different values of dendrite branching frequency. The lower boundary (blue in Figure 3C) of the collection of these curves defines the optimal performance of this combination of branching strategies (Strategy 1 for axons and Strategy 1 for dendrites). This performance boundary depicts the minimal number of branches that are required to establish the connectivity of given level of precision after a given length of physical time.

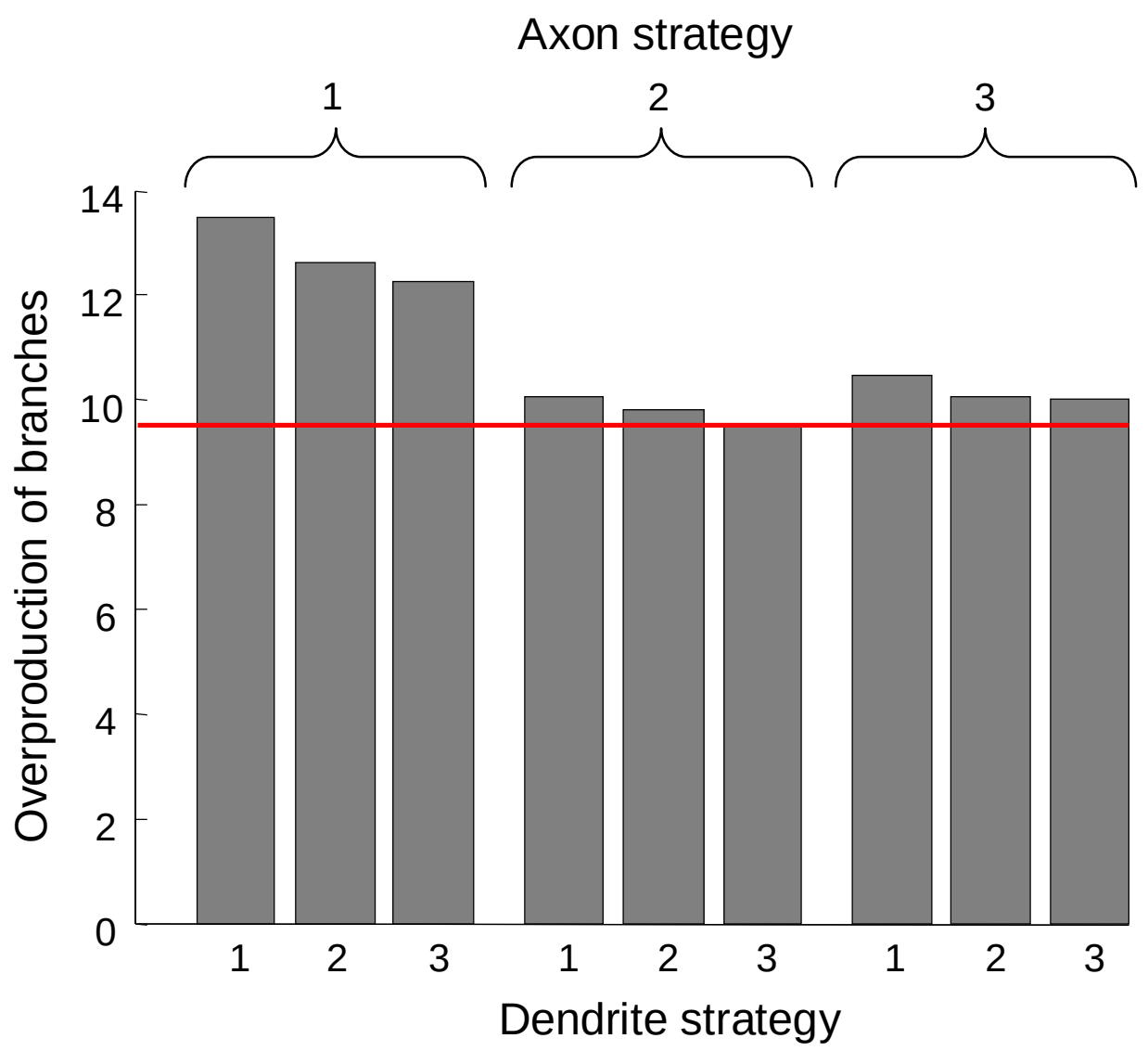

Figure 5: The average overproduction of branches for 9 combinations of branching strategy used by axons and dendrites. The overproduction of branches is the ratio of the number of branches formed to the minimal number of branches required to establish the topographic map. We then average this value over the performance boundaries from Figure 4. The most effective is the combination of branching Strategy 2 for axons and Strategy 3 for dendrites (level shown by red line).

To compare the efficiency of different branching strategies, we obtained the performance boundaries for all 9 combinations of strategies used by axons and dendrites; i.e., strategies 1 through 3 for axons and 1 through 3 for dendrites (Figure 4). The combination of branching strategies with the lowest boundary allows for the most effective formation of circuitry.

One of the findings evident from Figure 4 is that both synaptotropic strategies (activity-dependent and -independent) generally outperform the synapse-independent strategy. Thus, if both axons and dendrites implement synapse-independent branching (Figure 4, top left panel), the performance boundary represents the worst solution. This is because the performance boundary for this case is higher than all eight other performance boundaries. The same conclusion follows from examining the number of branches averaged along the performance boundary (Figure 5). The three bars on the left, representing the synapse-independent strategy implemented by axons, are higher than all others, reflecting the inefficiency of the synapse-independent branching rule. A similar conclusion is reached comparing the dendritic strategies (Figure 5). Therefore, for both axons and dendrites, synaptotropic branching improves the performance of the search algorithm over synapse-independent rules. This finding suggests a functional role for the spatial correlations between the branch points and synapses observed among axons (Alsina et al., 2001; Meyer and Smith, 2006). According to our results, the increased probability to form a branch point at an existing synapse (synaptotropic branching) allows for the establishment of required connectivity using fewer transient branches.

**The optimal branching rules are different for axons and dendrites**

What is the optimal synaptotropic branching strategy? According to our results (Figures 4 and 5), the most efficient combination of branching rules is achieved when axons implement Strategy 2 (synaptotropic activity-independent), while dendrites implement Strategy 3 (synaptotropic activity-dependent). The performance boundary for this combination of branching rules (red in Figure 4) is lower than all other eight curves. This implies that the optimal branching rules are different for axons and dendrites. To minimize the total amount of material spent, axons branch in the vicinity of existing synapses. But the frequency of such branching does not depend upon the correlations in patterned pre- and post-synaptic activity. At the same time, optimal dendritic branching is achieved when new branches are formed more likely in the vicinity of synapses with higher levels of correlated activity.

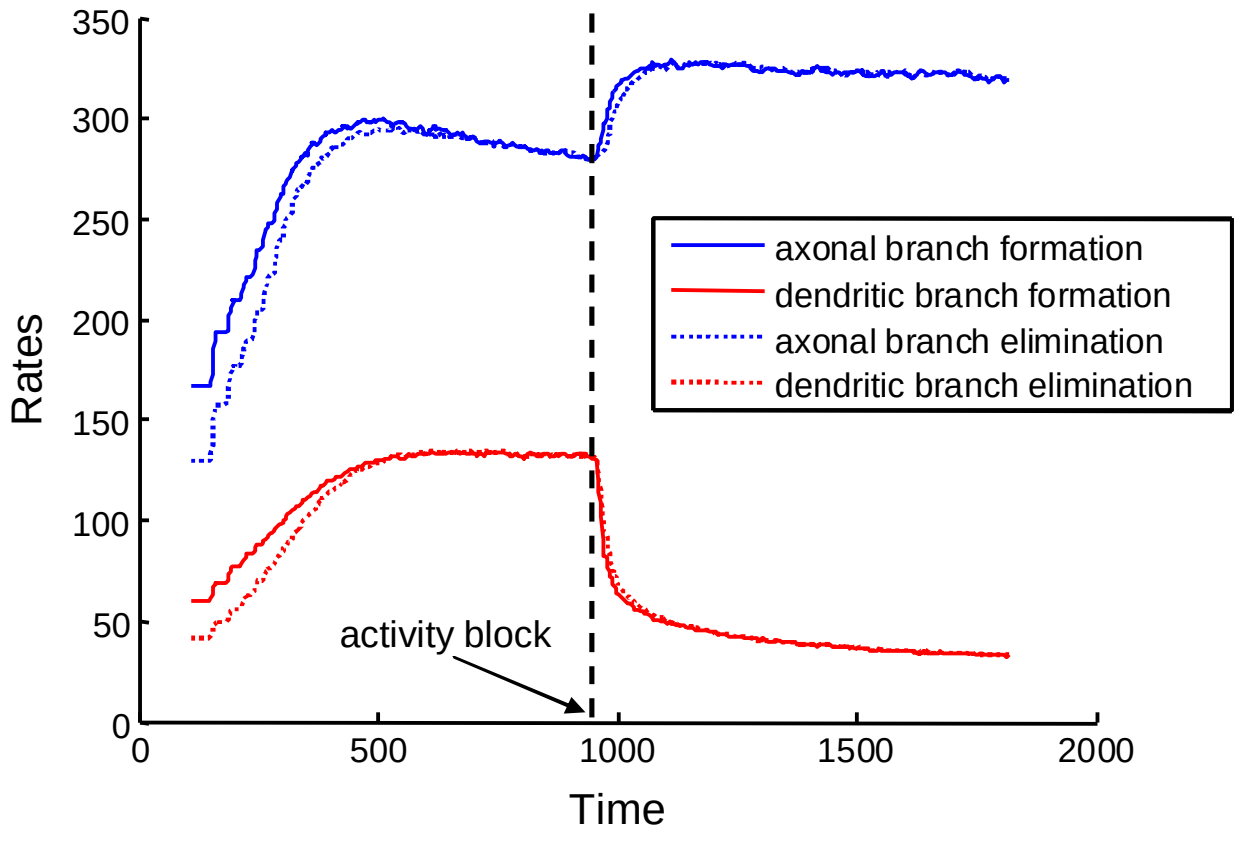

Figure 6: Rates of branch additions and retractions before and after activity block. The time dependence of axonal (blue) and dendritic (red) dynamics is shown for when axons implement the activity-independent synaptotropic branching strategy and dendrites use the activity-dependent branching strategy. This combination of branching strategies is optimal. The rates of branch additions (solid curves) and retractions (dotted curves) increase for axons and decrease for dendrites after the activity strength is set at zero (arrow) during simulation. In this simulation, the costs of dendritic and axonal branch lengths are unequal, so as to produce shorter dendrites.

If axons and dendrites implement different branching strategies, they can react differently to activity blockade. Consequently, our findings could explain the differences in the reaction to the blockade of NMDA receptors observed in developing retinotectal projections of *Xenopus laevis* (Rajan et al., 1999). To mimic the blockade of NMDA receptors in the model, we set the activity level to zero during simulations. We used the optimal combination of synaptotropic branching strategies (Strategy 2 for axons and Strategy 3 for dendrites). We observed that, for axons, both the rate of addition and retraction of branches increase after activity blockade (Figure 6). This is because, while the frequency of axonal branching does not change at any location on the arbor if axons implement activity-independent branching rules, the area occupied by the axonal arbor increases due to the loss of map precision induced by the activity block. Hence, larger arbors produce an increased rate of branch turnover.

At the same time, the rates of formation and elimination of dendritic branches are decreased after the levels of activity are reduced. This is a consequence of the activity-dependent branching rule (Strategy 3) implemented by dendrites, because the frequency of branching in the vicinity of the synapses is reduced. Therefore, in our model, the behavior of axons and dendrites is different, due to the differences in the optimal branching strategies. The experimentally-observed asymmetry in the reaction of axons and dendrites to NMDA receptor blockade could be a manifestation of different branching strategies being implemented by axons and dendrites in developing brain.

## Discussion

During neural development, axons solve the problem of locating the dendrites of appropriate cells and creating synapses with them. Finding appropriate synaptic partners occurs in the presence of other axons and dendrites, and is influenced by several factors, such as molecular labels and correlations in electric activity. How can precise connectivity be formed in the developing brain under the constraints of limited resources, like time and material? It is common in the computer sciences to benchmark different algorithms based upon the number of steps that they require to solve particular problems. The algorithm that solves a given problem with the smallest number of iterations usually is implemented. In this study, we benchmarked various algorithms for axonal and dendritic branching, and derived the branching rule that solves the problem of forming connectivity with the smallest number of steps. We assumed that the elementary step in the development of brain circuitry is the formation or elimination of an axonal or dendritic branch. We, thus, compared different branching rules, in terms of the total number of branches needed to form target circuitry. We assumed that the search strategy that allows for the location of targets using the fewest transient branches is implemented in the developing brain, in accordance with the wiring optimization argument (Chklovskii and Koulakov, 2004).

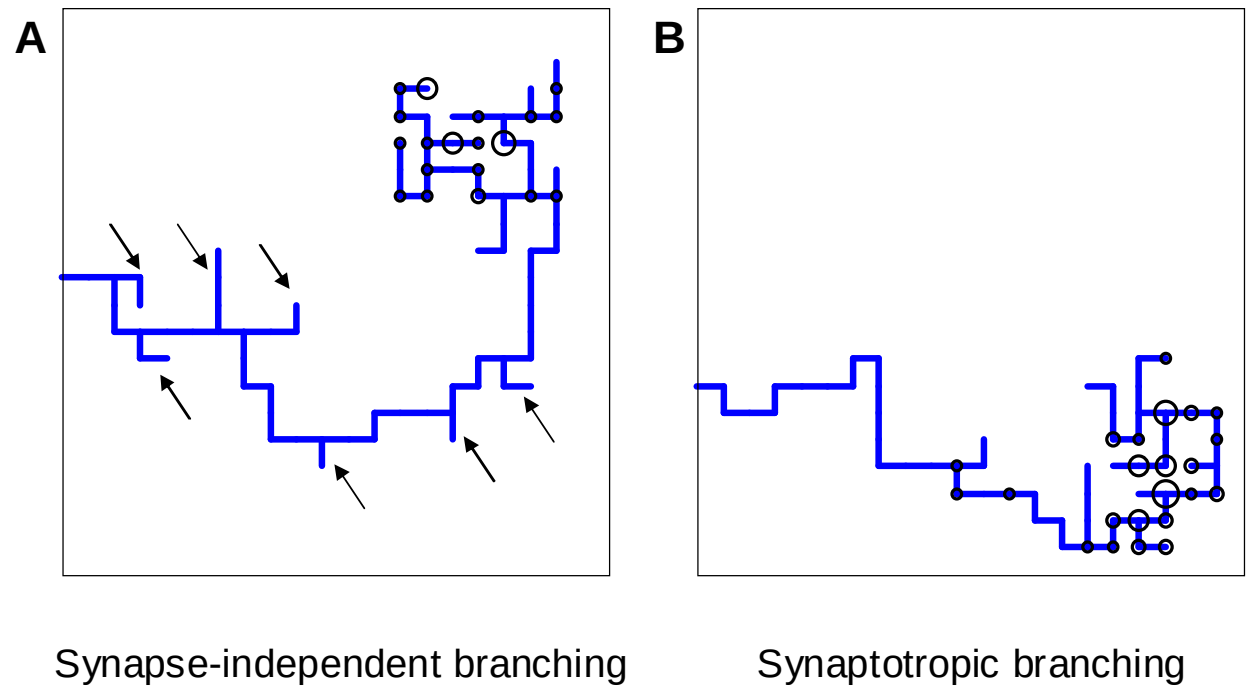

Figure 7: Comparing synapse-independent branching and synaptotropic branching for axons. (A) If the branching is synapse-independent, new branches are formed along all segments of the arbor, both proximal and distal to the correct TZ. (B) In the case of synaptotropic branching, new branches are formed only along the arboreal segment closest to the correct TZ, and there are no erroneous branches formed along segments that are far away from its future TZ (marked by arrows in A). Thus, synaptotropic branching allows for axons to reach the correct TZ using fewer branches.

We centered our studies on the role of synapses in the development of connectivity. In the developing retinotectal projection, synapses are formed and eliminated, as axons (Alsina et al., 2001; Meyer and Smith, 2006; Ruthazer et al., 2006) and dendrites (Niell et al., 2004; Haas et al., 2006; Sanchez et al., 2006) refine their connectivity. The role of synapses in this process may be diverse: they stabilize existing axon branches (Meyer and Smith, 2006) in a way that is dependent upon synaptic maturation (Ruthazer et al., 2006) and may contribute to the process of forming new branches (Meyer and Smith, 2006). The latter possibility is highlighted by strong correlations between the locations of synaptic puncta and the branch points observed for both axons and dendrites (Alsina et al., 2001). The effect of synapses on branch formation and elimination sets the basis for the synaptotropic hypothesis, according to which the formation of axonal and dendritic arbors is instructed by synapses. Here, we investigated the functional significance of the instructive role of synapses in the formation of new branches. To this end, we compared the branching rule that does not take into account the location of synapses (synapse-independent) with the synaptotropic branching rules. The latter make forming a new branch at the location of existing synapse more likely. We found that the synaptotropic branching rule allows for the formation of target connectivity using fewer erroneous branches (fewer steps). Consequently, our study implies that the functional significance of the observed correlations between branch points and synapses (Alsina et al., 2001) are the result of a frugal developmental mechanism.

We illustrate the advantage of synaptotropic branching rules for axons in Figure 7. The axonal arbor has segments proximal to its correct termination zone (TZ) and segments distal to its TZ. The formation of new branches along the proximal segments of the arbor contributes to arbor growth towards its TZ, while formation of new branches along distal segments is a waste of material. How can axons distinguish between the proximal and distal regions of the arbor? The transient synapses are located on the segments of the arbor closest to the correct TZ, because they are made with more appropriate dendritic partners and, thereby, are more stable. As a result, synaptotropic branching allows for avoidance of the formation of erroneous branches on distal parts of the arbor, and for establishing spatial overlap between appropriate pairs of axonal and dendritic arbors using fewer steps.

We have further investigated the possible role of correlated electric activity on the synaptotropic branching rules. We compared the activity-independent synaptotropic branching strategy when new branches are formed with the same probability in the vicinity of all existing synapses versus activity-dependent synaptotropic branching when the branches are preferentially formed at the synapses with high correlations between the activity of pre- and post-synaptic cells. We found that there is a slight *decrease* in the total number of branches used, if dendrites but not axons implement activity-dependent synaptotropic branching. These results suggest that axons and dendrites have different optimal branching strategies.

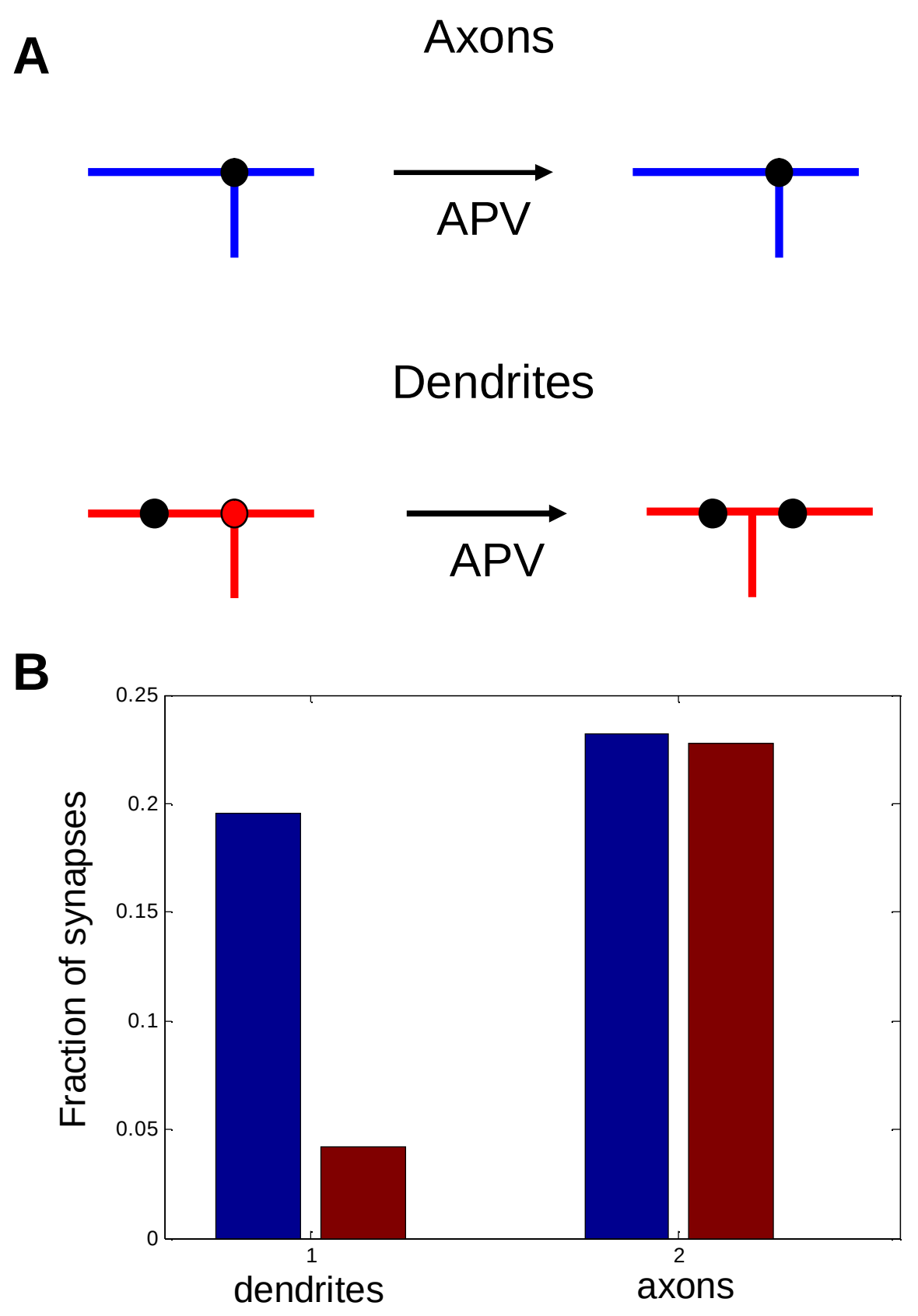

Figure 8: The role of activity block on the spatial correlations between branch points and synapses. (A) Schematic representation of the predictive capacity of the model. If axons implement an activity-independent synaptotropic branching strategy and dendrites implement an activity-dependent synaptotropic branching strategy, the consequences of activity blockade are different for axons and dendrites. The spatial correlations between synapses and branch points remain unchanged for axons, and are reduced for dendrites after an NMDAR antagonist is applied. (B) The fraction of synapses located at branch points before (blue) and after (red) activity block in the simulation, when axons and dendrites implement optimal branching strategies.

What is the origin of asymmetry in the branching rules between axons and dendrites? Axons and dendrites solve the same search problem during the formation of connectivity. We argue that the origin of the asymmetry between axons and dendrites is in the difference that exists in the initial conditions of the search problem they face. When axons of retinal ganglion cells enter the optic tectum, they lack spatial order. In contrast, dendritic arbors originate from somas of tectal cells that are topographically arranged in the target. Axons use the topographic arrangement of dendrites to locate the correct termination zone efficiently, by implementing a synaptotropic branching strategy (Strategy 2) which instructs axons *where* to branch, as discussed above.

Dendrites, on the other hand, do not grow towards distal TZ, because they are approached by their appropriate axonal partners from uncertain directions. Instead, dendritic branching increases the cross-section between overlapping arbors of axons and dendrites. This enhances the probability of synapse formation between correct partners, and results in the faster refinement of connectivity. This refinement takes place after the initial overlap between the correct axonal and dendritic partners is established. Correlations in the activity between pre- and postsynaptic cells increase over time as the topographic map is formed. Thus, the level of these correlations may signal the dendrites when crude topography is established and their ultimate axonal partners reach the correct TZ. Implementing the activity-dependent branching strategy (Strategy 3) lets dendrites increase the formation of new branches, only when they are contacted by appropriate axons, thereby reducing the number of transient branches formed. In other words, activity-dependent branching strategy instructs dendrites *when* to form new branches.

The hypothesis that axons and dendrites use different branching strategies is consistent with existing experiments on the blockade of NMDA receptors (Rajan et al., 1999). Our modeling shows that axons accelerate the formation and elimination of new branches, while the branching of dendrites slows down under conditions of reduced correlated activity (Figure 6). The latter finding is a direct result of the instructive role of activity in dendritic, but not axonal branching. In our model, acceleration of axonal branching is a result of removing the activity-dependent stabilization of synapses that exists in the condition of NMDA receptor blockade, which ultimately leads to more dynamic axons. Our results permit us to interpret the asymmetry that transpires in the changes occurring in the dynamics of branch formation, under the conditions of NMDA receptor blockade (Rajan et al., 1999).

We now propose ways in which this asymmetry in branching strategy can be further tested experimentally. We suggest that, if an NMDAR antagonist is applied to developing optic tectum, the spatial correlations between *axonal* branch points and synapses should remain the same as in the control case. In contrast, the correlations between the locations of *dendritic* branch points and synapses should be reduced after NMDAR blockade. We illustrate this prediction in Figure 8, where we measure the fraction of synapses that are located in the vicinity of axonal and dendritic branch points before and after activity block in the model. Such observations recently were made in *Xenopus* for both axons (Meyer and Smith, 2006; Ruthazer et al., 2006) and dendrites (Niell et al., 2004; Sanchez et al., 2006) without the application of NMDAR antagonists.

We propose the functional role of axonal and dendritic branching from a developmental point of view. With this approach, branching is required to speed up the developmental process. Acceleration in the location of correct targets due to branching is accomplished via the use of a parallel search algorithm. Another possibility is that branching is required to optimize the functionality of the mature circuit; for example, to improve the signal transmission properties of the network (Poirazi et al., 2003; Wen and Chklovskii, 2005). One way to confirm the relevance of developmental optimality of branching is to analyze the branch dynamics and arbor morphology in different brain regions and across species. It is plausible that optimal branching strategies differ for the development of circuitry in different regions of the brain and/or in different species. These differences may reflect specifics in local circuitry or initial conditions. Thus, comparing branch dynamics and arbor morphologies in different systems may reveal the importance of the optimality of the search algorithms used in developing brain.

In conclusion, we studied computationally-different branching rules for axons and dendrites within a developing retinotectal projection. Our studies suggest that branching serves to accelerate the formation of neuronal circuitry through the use of the parallel search of targets. We argue that the observed abundance of synapses on branch points for both axons and dendrites serves to minimize the number of erroneous transient branches. We also explain the asymmetry that is observed experimentally in the reaction to NMDA receptor blockade between axons and dendrites. We suggest that this asymmetry stems from the branching of dendrites, but not axons, being directly instructed by correlations in electric activity. Finally, we propose experimental tests that could verify that optimal branching rules, indeed, are being implemented in developing brain.

## Methods

We propose a mathematical description of the dynamics of axonal and dendritic arbors, using the theoretical model of stochastic growth. In this model, new branches are created, eliminated, elongated and retracted randomly, with probabilities dependent upon how the energy of the system changes after a segment of a branch is added or removed. The neural connections are formed in the model by creating synapses between the branches of axons and dendrites spatially located in the same volume. Similar to branches, the synapses can be maintained or retracted later, in a stochastic manner, depending upon the energy change that transpires after a synapse is removed. We simulate the developmental process using the *Metropolis Monte Carlo* algorithm. At each *Monte Carlo* step, one of the six changes is attempted: formation, elimination, extension or retraction of branches, or creation or elimination of synapses. The acceptance probabilities depend upon the change in the energy function that occurs during these processes, $\Delta E$, and are given by

$$p = 1/2 + 1/2 \tanh(-2\Delta E).$$

As a result, the system performs the stochastic minimization of its free energy. The exact form of the energy function defines both the dynamics of the arbors and the final connectivity configuration (Tsigankov and Koulakov, 2006).

In our model, the energy function incorporates the affinity that exists between connected cells and the material cost of the arbors. It contains additive contributions from axonal and dendritic branches and synaptic connections:

$$E = E_{ax.arb} + E_{den.arb} + E_{syn}.$$

The contribution from arbors to the energy function is positive, meaning that there is a cost associated with the formation of branches. We also suggest that the synaptic contribution is negative, reflecting the tendency of neurons to form synapses. This contribution is different in magnitude for every synapse, and depends upon interactions between the connected cells. Combined together, these contributions support the synaptotropic hypothesis (Hua and Smith, 2004; Meyer and Smith, 2006), since the cost of a branch bearing a synapse is reduced, and such a branch is more stable than a branch without synapses.

The synaptic part of the energy function depends upon the layout of connectivity between axons and dendrites, given by the weight matrix $W_{ij}$. We previously have studied the form of the synaptic energy function for the system of point-like axons and dendrites (Tsigankov and Koulakov, 2006). Here, we reformulate it for the system of axons and dendrites with spatially distributed arbors that have multiple synaptic connections. There are three additive terms in the model, representing different biological contributions:

$$E_{syn} = E_{chem} + E_{act} + E_{comp}$$

The chemoaffinity term depends upon the interactions between the chemical labels expressed on axons and dendrites. For the retinocollicular system, this is given by the expression levels of EphA and EphB receptors on axons, and of ephrinA and ephrinB ligands on dendrites:

$$E_{chem} = \sum_{\alpha\beta} M_{\alpha\beta} \sum_{ij} L_i^{\alpha} W_{ij} R_j^{\beta}.$$

Here, indices $\alpha$ and $\beta$ denote the chemical labels; the matrix $M_{\alpha\beta}$ defines the affinities for receptor/ligand pairs; and $L_i^{\alpha}$, $R_j^{\beta}$ are the concentrations of ligand $\alpha$ and receptor $\beta$ on the $i^{th}$ dendrite and the $j^{th}$ axon, respectively. Throughout the paper, we have adopted the simplest description, where we distinguish only two types of receptor and ligand expressed in the gradients in perpendicular directions in both the target and retina; for details see (Koulakov and Tsigankov, 2004; Tsigankov and Koulakov, 2006).

The activity-dependent term is obtained from the Hebbian learning rule and has the form:

$$E_{act} = -\frac{1}{2} \sum_{ij} W_{ij} D_{ij}.$$

Here, $D_{ij}$ is the correlation of electrical activity between dendrite $i$ and axon $j$. It is computed from the correlations of activity between axons that are projecting to a given dendrite, using the following expression:

$$D_{ij} = \sum_{lm} W_{lm} U_{il} C_{jm}.$$

$C_{jm}$ is the correlation of activity between axons $j$ and $m$, and $U_{il}$ is the strength of the Hebbian interaction between dendrites $i$ and $l$. Both these functions are presumed to depend only upon the spatial separation that exists between the origins of axon $j$ and $m$ and dendrite $i$ and $l$, respectively (Tsigankov and Koulakov, 2006):

$$C_{jm} = \exp\left(-\left|\vec{r}_j - \vec{r}_m\right| / a\right),$$

$$U_{il} = \gamma \exp\left(-\left|\vec{r}_i - \vec{r}_l\right|^2 / 2b^2\right).$$

The last term in the synaptic part of the energy function describes axonal and dendritic competition and ensures the tendency of neurons to form synapses. This term is negative and depends upon the number of synapses made by each neuron, as proposed for the system of neuromuscular junctions (Barber and Lichtman, 1999). If the energy gain decreases with an increase in the total number of synapses per cell, then cells with fewer synapses have a competitive advantage to form new synapses. As a result, every axon and every dendrite maintains approximately the same number of synapses. In our model, we used the following form of energy contribution with this property that has the least number of parameters:

$$E_{comp} = -b_a \sum_j \left( \sum_i W_{ij} \right)^{1/2} + b_d \sum_i \left( \sum_j W_{ij} \right)^2 .$$

The sums in the brackets give the number of synapses made by axons and dendrites respectively; $b_a > 0$ and $b_d > 0$ are the constants defining the overall strength of axonal and dendritic competition.

The arbor parts of the energy function that describe the costs for axonal and dendritic branching are given by

$$E_{ax.arb.} = \sum_{ax.br} \mu_l^a l + \sum_{ax.bp} \mu_{bp}^a ,$$

$$E_{den.arb.} = \sum_{den.br} \mu_l^d l + \sum_{den.bp} \mu_{bp}^d ,$$

where the first sum over axonal and dendritic branches yields the cost for the branch with length *l*, and the second sum represents the additional cost for the formation of the branch points. We assume that the costs of the branches per unit length $\mu_l^a$ and $\mu_l^d$ are constant and are taken to be the same throughout the paper, so as to ensure symmetry between axons and dendrites.

In contrast, the costs of branch points $\mu_{bp}^a$ and $\mu_{bp}^d$ can vary and have different forms for different branching strategies. If axons or dendrites use synapse-independent branching Strategy 1, we have a constant branching cost for all axons or dendrites

$$\mu_{bp} = \mu_0 = const .$$

For activity-independent synaptotropic branching Strategy 2, we use

$$\mu_{bp} = \mu_0 / n_s ,$$

where $n_s$ is the number of synapses on the arbor in the vicinity of a branch point.

Finally, for activity-dependent synaptotropic branching Strategy 3, the cost has the form

$$\mu_{bp} = \frac{\mu_0}{\sum_s D_{i(s)j(s)}} .$$

Here, the sum is taken over synapses made on branches at the location of the branch point, and $D_{ij}$ is the correlation of the electrical activity between dendrite *i* and axon *j* connected with these synapses. In this description, for every branching strategy used, we can vary the overall amount of branching by changing the single parameter $\mu_0$.

## References

Alsina B, Vu T, Cohen-Cory S (2001) Visualizing synapse formation in arborizing optic axons in vivo: dynamics and modulation by BDNF. Nat Neurosci 4:1093-1101.

Barber MJ, Lichtman JW (1999) Activity-driven synapse elimination leads paradoxically to domination by inactive neurons. J Neurosci 19:9975-9985.

Chklovskii DB, Koulakov AA (2004) Maps in the brain: what can we learn from them? Annu Rev Neurosci 27:369-392.

Flanagan JG, Vanderhaeghen P (1998) The ephrins and Eph receptors in neural development. Annu Rev Neurosci 21:309-345.

Fraser SE, Perkel DH (1990) Competitive and positional cues in the patterning of nerve connections. J Neurobiol 21:51-72.

Haas K, Li J, Cline HT (2006) AMPA receptors regulate experience-dependent dendritic arbor growth in vivo. Proceedings of the National Academy of Sciences of the United States of America 103:12127-12131.

Hamos JE, Van Horn SC, Raczkowski D, Sherman SM (1987) Synaptic circuits involving an individual retinogeniculate axon in the cat. J Comp Neurol 259:165-192.

Hua JY, Smith SJ (2004) Neural activity and the dynamics of central nervous system development. Nat Neurosci 7:327-332.

Hua JY, Smear MC, Baier H, Smith SJ (2005) Regulation of axon growth in vivo by activity-based competition. Nature 434:1022-1026.

Koulakov AA, Tsigankov DN (2004) A stochastic model for retinocollicular map development. BMC Neurosci 5:30.

Lemke G, Reber M (2005) Retinotectal Mapping: New Insights from Molecular Genetics. Annu Rev Cell Dev Biol.

McLaughlin T, O'Leary DD (2005) Molecular gradients and development of retinotopic maps. Annu Rev Neurosci 28:327-355.

McLaughlin T, Torborg CL, Feller MB, O'Leary DD (2003) Retinotopic map refinement requires spontaneous retinal waves during a brief critical period of development. Neuron 40:1147-1160.

Meyer MP, Smith SJ (2006) Evidence from in vivo imaging that synaptogenesis guides the growth and branching of axonal arbors by two distinct mechanisms. The Journal of neuroscience : the official journal of the Society for Neuroscience 26:3604-3614.

Niell CM, Meyer MP, Smith SJ (2004) In vivo imaging of synapse formation on a growing dendritic arbor. Nature neuroscience 7:254-260.

Pfeiffenberger C, Cutforth T, Woods G, Yamada J, Renteria RC, Copenhagen DR, Flanagan JG, Feldheim DA (2005) Ephrin-As and neural activity are required for eye-specific patterning during retinogeniculate mapping. Nat Neurosci 8:1022-1027.

Poirazi P, Brannon T, Mel BW (2003) Arithmetic of subthreshold synaptic summation in a model CA1 pyramidal cell. Neuron 37:977-987.

Rajan I, Witte S, Cline HT (1999) NMDA receptor activity stabilizes presynaptic retinotectal axons and postsynaptic optic tectal cell dendrites in vivo. J Neurobiol 38:357-368.

Ruthazer ES, Cline HT (2004) Insights into activity-dependent map formation from the retinotectal system: a middle-of-the-brain perspective. Journal of neurobiology 59:134-146.

Ruthazer ES, Akerman CJ, Cline HT (2003) Control of axon branch dynamics by correlated activity in vivo. Science (New York, NY) 301:66-70.

Ruthazer ES, Li J, Cline HT (2006) Stabilization of axon branch dynamics by synaptic maturation. The Journal of neuroscience : the official journal of the Society for Neuroscience 26:3594-3603.

Sanchez AL, Matthews BJ, Meynard MM, Hu B, Javed S, Cohen Cory S (2006) BDNF increases synapse density in dendrites of developing tectal neurons in vivo. Development (Cambridge, England) 133:2477-2486.

Sin WC, Haas K, Ruthazer ES, Cline HT (2002) Dendrite growth increased by visual activity requires NMDA receptor and Rho GTPases. Nature 419:475-480.

Stepanyants A, Chklovskii DB (2005) Neurogeometry and potential synaptic connectivity. Trends in neurosciences 28:387-394.

Tsigankov D, Koulakov A (2006) A unifying model for activity-dependent and activity-independent mechanisms predicts complete structure of topographic maps in ephrin-A deficient mice. . J Computational Neuroscience:In Press.

Wen Q, Chklovskii DB (2005) Segregation of the brain into gray and white matter: a design minimizing conduction delays. PLoS Computational Biology 1:e78.